\documentclass[prd,aps,floats,preprintnumbers,preprint,floatfix]{revtex4}

\usepackage{graphicx, epsfig}
 
\newcommand{\ee}{\end{equation}}
\newcommand{\bea}{\begin{eqnarray}}
\newcommand{\eea}{\end{eqnarray}}

\begin{document}
 
\thispagestyle{empty}
\begin{flushright}
\parbox[t]{1.5in}{MAD-TH-03-2 \\
MAD-PH-03-1359}
\end{flushright}

\vspace*{0.5in}

\begin{center}
{\Large \sc D-Matter}

\vspace*{0.5in} 
{\large Gary Shiu and Lian-Tao Wang}\\[.3in]
{\em Department of Physics,
  University of Wisconsin,
Madison, WI 53706, USA}\\[0.5in]
\end{center}

\begin{center}
{\bf
Abstract}
\end{center}

We study the properties and
phenomenology of particle-like states originating from D-branes
whose spatial dimensions are all compactified. They are non-perturbative states
in string theory and we refer to them as {\it D-matter}. 
In contrast to other non-perturbative objects such as 't Hooft-Polyakov
monopoles, 
D-matter states could have 
perturbative couplings among themselves and
with ordinary matter. 
The lightest D-particle (LDP) could be stable because it is the
lightest state carrying certain (integer or discrete) quantum numbers.
Depending on the string scale, they could be  
cold dark matter candidates with
properties similar to that of wimps or wimpzillas.
The spectrum of excited states of 
D-matter exhibits an interesting pattern which could be distinguished
from that of 
Kaluza-Klein modes, winding states, and string resonances.
We speculate about possible signatures of D-matter from ultra-high
energy cosmic rays and colliders. 

\vfill

\hrulefill\hspace*{4in}

{\footnotesize
Email addresses: shiu@physics.wisc.edu, liantaow@pheno.physics.wisc.edu.}

\newpage

\section{Introduction}

In addition to their pivotal role in elucidating non-perturbative
aspects of string theory, in recent years D-branes have become
ubiquitous
in particle phenomenology and cosmology.
In the brane world scenario \cite{HW,ADD,TeV,Lykken,Ovrut,RS}, 
the Standard Model gauge bosons
and matter fields are localized on some space-filling branes whereas
only gravity  
and closed string modes (such as the radions) can propagate in the bulk, 
thus offering the possibility of a lower fundamental energy scale.
It is worth noting that in this
framework
the D-branes are part of the string theory vacuum.
The Standard Model particles and the bulk modes are excitations 
above the vacuum
described by {\it perturbative} string states in the background of D-branes. 
Therefore, 
even though the branes themselves are non-perturbative objects,
the physics of the brane world scenario can be studied using standard
techniques of string perturbation theory.

In this paper, we explore yet a different phenomenological feature of D-branes.
Unlike the background D-branes in
the context of brane world, the D-branes we consider are not space-filling
but rather behave as point particles in four dimensions.
These particle-like states arise when all spatial dimensions of $Dp$-branes
(irrespective of $p$) are wrapped around the compact space. 
We refer to them collectively
as {\it D-matter}.
As we will discuss in more detail, these D-matter states appear quite
naturally in four-dimensional string models. They are non-perturbative objects 
with mass $m \sim M_s/g_s$ where $g_s$ is the string coupling 
and so at weak coupling they are heavier than the 
perturbative string states.
However, in order to produce the observed gauge and gravitational couplings, 
$g_s$ is not arbitrarily small but is typically of order one.
Hence, the D-matter states can be sufficiently light to be phenomenologically 
and cosmologically relevant.
The D-matter states are dynamical degrees of freedom and so, 
in contrast to the background D-branes, we will treat them as
excitations above the vacuum.

The stability of D-branes is due to 
the charges that they carry. 
Depending on their types (BPS or non-BPS), D-branes could 
carry integral or torsion (discrete) charges.
In fact, it is because of the torsion charges that the complete
spectrum of D-branes  should be classified by 
K-theory \cite{Moore,Witten,Horava}.
Among the D-matter states, 
the lightest D-particle (LDP) is stable because it is the lightest
state carrying its specific charges.
Therefore, just like the lightest supersymmetric partners 
(LSPs) in supersymmetric models, the LDPs are possible candidates for
cold dark matter. As we will show, an important difference between the
D-matter states and other non-perturbative objects (such as magnetic
monopoles \footnote{Magnetic monopoles are common in string models
and their
cosmological implications have been explored
\cite{Witten:2002wb}. As we will discuss, there are similarities and differences
between D-matter and magnetic
monopoles.}) is that they could have {\bf {\it perturbative} }couplings. 
Hence, the LDPs are weakly interacting and so they could be candidates
for wimps or wimpzillas depending on the string scale.
We also comment on the case where the D-matter states are 
unstable (i.e., the spectrum of
stable D-branes does not contain four-dimensional particle-like states).
Although they are not long-lived to be cosmologically relevant, they could
be produced at colliders and give rise to interesting signatures.

In addition to D-matter which are particle-like states (from a four-dimensional perspective), 
other types of stable defects 
(such as cosmic strings and domain walls)
could also exist in a string vacuum.
Given a specific string model, it is straightforward to deduce
the complete spectrum of such defects.
In this paper, we focus on the general properties of D-matter 
 and point out their relevance to
  phenomenology. 
We also stress that the cosmological constraints on the
various kinds of stable defects could provide important 
additional
criteria on the viability of a string model (and hence in
estimating the number of realistic string vacua \cite{Douglas:2003um}).
For example, we can rule out a string vacuum if there exist stable electrically charged
D-matter states.

This paper is organized as follows. In Section \ref{properties}, we
describe the  properties of D-matter including its stability, mass,
and interactions.   In Section \ref{phenomenology},
we discuss the phenomenological implications and some possible signatures 
of D-matter. Finally, we end with some discussions and conclusion in
Section \ref{conclusion}. Some details of
the four-point heterotic string amplitude which is dual
to the annihilation cross-section of D-matter states are relegated to the appendix.

\section{Properties of D-matter}\label{properties}

\subsection{Mass of D-matter}

D-matter has a very simple origin. A $Dp$-brane sweeps out a $p+1$
dimensional worldvolume.
From a four-dimensional perspective, 
a D-brane with all of its spatial dimensions compactified will sweep
out a worldline and behave like a point particle.

In this section, we will consider the constraints on the mass of 
D-matter. Since D-matter states arise from $Dp$-branes wrapping around
$p$ compactified dimensions, their masses are
given by:
\begin{equation}
m_D = \frac{M_s^{p+1} V_p}{g_s}~,
\end{equation}
where $V_p$ is the volume of the $p$-cycle (i.e., 
$p$-dimensional subspace of the internal manifold)
that the $Dp$-branes wrapped around. The D-matter states 
are non-perturbative objects whose masses are inversely proportional to $g_s$.
At weak coupling, they are heavier than the perturbative string states.
However,  $g_s$ cannot be arbitrarily small for otherwise the Planck mass $M_P$
is not finite. This can be seen from the relation:
\begin{equation}
M_P^2 = \frac{M_s^8 V_6}{g_s^2}
\end{equation}
where $V_6$ is the overall volume of the internal manifold.
Hence, $m_D$ is not much heavier than $M_P$.
As we shall see shortly, if the dimensions transverse to the $D_p$
brane have size $R_{\perp}$  larger than
the string scale $M_s R_{\perp} >>1$, the resulting D-matter state can
be much lighter than $M_P$. 

\begin{center}
\begin{figure}[h]
\centering
\epsfxsize=12cm
\epsfbox{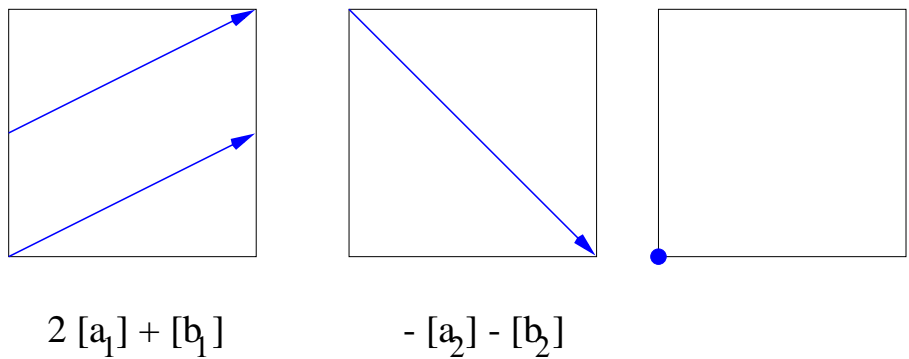}
\label{cycles}
\caption{A stable D2-brane wrapping around the 2-cycle $(2 [a_1] +
[b_1]) \times (-[a_2]-[b_2])$.}
\end{figure}
\end{center}

We can obtain a more quantitative estimate of $m_D$.
Although in general there is no direct relation between $V_p$ and $V_6$, 
for simple compactifications such as toroidal and orbifold backgrounds,
we can show that generically $M_s/g_s \lesssim m_D \lesssim M_P$.
For illustrative purpose, let us consider the internal manifold to be a product of 
tori $T^2 \times T^2 \times T^2$ (or its orbifold thereof) with sizes
$R_i$ where $i=1,\dots,3$, then
\begin{eqnarray}
M_P^2 &=& \frac{M_s^8 \prod_{i=1}^3 R_i^2}{g_s^2}~, \\
m_D &=& \frac{M_s^{p+1} V_p}{g_s} \sim \frac{M_s^{p+1} \prod_{j=1}^p L_j}{g_s},
\end{eqnarray}
where we assume that the p-cycle that the 
$Dp$-brane wrapped around is factorisable into a product of one-cycles
with lengths $L_j$ (which are related to $R_{i}$ by the wrapping
numbers). 
Let us denote
a canonical basis of one-cycles in each $T^2$ by $[a_i]$ and $[b_i]$ (where
$i=1,2,3$) respectively. As an example, suppose that the D-matter
state originates from a stable D2-brane wrapped around a
two-cycle $\left( n_1 [a_1] + m_1 [b_1] \right)\times \left(n_2 [a_2] + m_2 [b_2]\right)$ as shown in Fig.~2, then $p=2$, and
$L_i = \sqrt{n_i^2 + m_i^2} R_{i}$.
The wrapping numbers are typically of order one and so $L_i \sim R_i$.
Note that D-brane has a size of the order of
$M_s^{-1}$ and so if $R_i < M_s^{-1}$, 
it is more
appropriate to go to the T-dual picture:
\begin{equation}
R_i \rightarrow\frac{1}{M_s^2 R_i} \qquad g_s \rightarrow \frac{g_s}{M_s R_i}
\end{equation}
and the $Dp$ brane becomes a $D(p-1)$ or a $D(p+1)$ brane
depending on whether $R_i$ is along or transverse to the worldvolume of
the $Dp$ brane. Therefore, there is a
natural lower bound for the size of the compact dimension $R \geq M_s^{-1}$. 
As a result, the mass of a D-matter state is bounded below by
$M_s/g_s$. For the same reason, $m_D \lesssim M_P$ and furthermore
if some dimensions transverse to the
$Dp$ branes are large, the D-matter state can be much lighter than $M_P$.

The mass of D-matter depends on the value of $M_s$.
In the brane world scenario, $M_s$ is not tied to $M_P$ and since $g_s
\sim O(1)$, the D-matter mass can be anywhere
between the TeV scale and the Planck scale.
As we will discuss, the D-matter states could have some interesting
phenomenological consequences if they are sufficiently light.

\subsection{Stability of D-matter}\label{stability}

Let us denote a D-brane which has
$r$ Neumann directions in our usual four-dimensional spacetime
and $s$ Neumann directions in the compact space
as a $D(r,s)$-brane. In this notation, 
there exist stable D-matter states if the
spectrum of stable D-branes contains some $D(r,s)$-branes with $r=0$
(irrespective of $s$). 

The complete spectrum of stable D-branes in a given string theory
background is rather rich and in general non-trivial to work out.
This is because contrary to naive expectations, D-brane charges 
are not classified by cohomology but instead by K-theory
\cite{Moore,Witten,Horava}. 
In addition to the BPS D-branes, there are generically non-BPS D-branes
that are stable \cite{Sen,Witten,Bergman,Frau}. 
The existence of such stable non-BPS branes is further exemplified by
some concrete orbifold \cite{Gaberdiel,Stefanski:orbifold}
and orientifold \cite{Quiroz,Braun} models.

The spectrum of stable D-branes can be divided into the following types:
\begin{itemize}
\item  {\it BPS states} 
are stable 
because they carry 
Ramond-Ramond (RR) charges, which are conserved charges 
associated with gauge fields coming from the RR sectors 
of string theory. The RR charges
are integrally quantized, analogous to the winding number of
a monopole solution. As a result, the BPS state with the smallest unit of charge ($n=1$) is
stable. An example 
of a stable BPS brane is the
$D0$-brane in Type IIA string theory (which behaves as particles even before compactification).

\item {\it Stable non-BPS states} which can be further divided into the following types:
\begin{enumerate}
\item  Non-BPS branes which carry twisted RR charges,
i.e., they are charged under the RR gauge fields localized at
orbifold fixed points. 
These twisted RR charges are quantized, just
like the untwisted RR charges carried by BPS branes. Therefore, the lightest
state carrying some given twisted RR charges is stable. 
\item Torsion charged D-branes. They are non-BPS branes which are stable even though they do not carry any RR charges. Rather, they carry some torsion (discrete) charges. 
It is precisely
these torsion charges that K-theory differs from cohomology. 
An example of such torsion-charged D-branes is the
stable D-particle in Type I string theory \cite{Sen}. Although $D0$-brane is unstable in Type IIB string theory, the tachyon which
signals the instability is projected out by the orientifold projection. Hence,
in Type I string theory (which can be obtained from Type IIB 
theory by an orientifold projection), the D-particle is stable and
in fact carries a ${\bf Z}_2$ torsion charge.

\end{enumerate}
\end{itemize}

The lightest D-matter states arising from some stable D-branes wrapping around
the compactified dimensions are stable because they are the lightest states
carrying certain (integer or discrete) quantum numbers.
In what follows,
we will refer to them as LDPs (lightest D-particles).

\subsection{Interactions of D-matter}\label{interaction}

Let us discuss how the D-matter states interact with
each other and with the Standard Model particles. 
To be concrete, we consider a specific case where the
D-matter states are the stable D-particles in Type I string theory and
the Standard Model is embedded on a set of D9-branes. 
Although we focus our analysis on this particular case for illustrative purposes,
we expect the properties of their interactions discussed below
are applicable to a general class of D-matter states.
The general rules for computing amplitudes involving the stable
Type I D-particles have been given in \cite{Sen-interactions,Witten}
and such rules 
have been applied to studying their gauge and gravitational
interactions in \cite{Gallot}. In the brane world scenario, 
the gauge bosons and the matter fields of the Standard Model
can in general be localized differently in the extra dimensions. For
example, in the context of intersecting brane world
\cite{CSU1,CSU2,CSU3,CIS,Berkooz,Berlin1,Sagnotti,Madrid1,Berlin2,Madrid2,Berlin3,Honecker,Bachas,Angelantonj},
the gauge bosons are localized on several different stacks of $Dp'$-branes
with $p' \geq 3$ whereas chiral matter fields are localized at the
brane intersections.  For simplicity, however, let us consider the
case where all Standard Model fields 
are localized on a single stack of $Dp'$-branes, keeping in mind that
chiral fermions can still arise in this scenario when the D-branes are
located at some singularities of the internal manifold.

The worldvolume of the $Dp'$-branes (where $p'=9$ in this particular case, in general $p' \geq 3$)
contains our usual four-dimensional space-time
and in general also some compactified dimensions. The D-matter state
propagates in the background of the higher-dimensional 
$Dp'$-branes where the Standard Model
is localized. 
It is worth emphasizing that in this "branes-within-branes" picture,
the D-matter states are not part of the background but instead its
{\it excitations}. Since $g_s$ is non-zero, their masses are finite  
and so they are dynamical degrees of freedom propagating
in some background D-branes.

In this simple setup, the Standard Model fields are open
string excitations with both endpoints of the open strings attached to
the same D$p'$-brane. These open string states are 
denoted by $C_{p'p'}$. There are also open strings
that stretched between the propagating D-matter state 
and the background $Dp'$-brane and we
denote these states by $C_{0 p'}$. As illustrated in Fig. 2, 
the $C_{0 p'}$ states couple to the Standard
Model fields in the $C_{p'p'}$ sector. Hence, 
there is an effective coupling of D-matter
to the gauge bosons (and other Standard Model fields) via the interactions
between perturbative open string states.
The D-matter states are charged under the gauge groups localized
on the $Dp'$-branes. Therefore, they 
couple 
with matter fields on the branes through gauge interactions.

\begin{center}
\begin{figure}[h]
\centering
\epsfxsize=12cm
\epsfbox{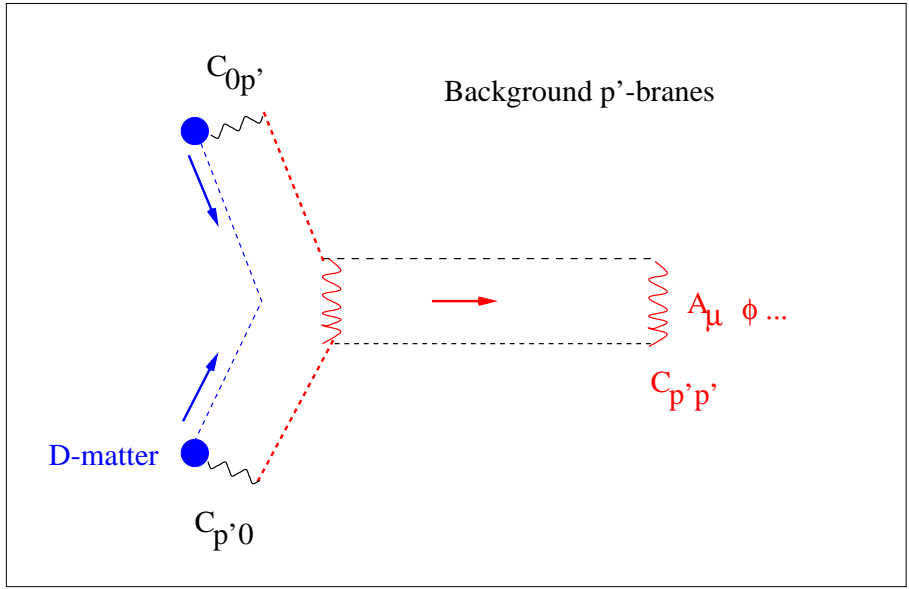}
\caption{D-matter can interact via perturbative string states.}
\end{figure}
\end{center}

Such D-matter states have gauge couplings and so they can
interact with each other by the exchange of gauge bosons. An
important aspect of the gauge (or matter) coupling is that it is
perturbative. At lowest order, the interaction of such D-matter with the
Standard Model fields is  
given by the disk diagram in Fig. 3 where $V_{0p'}$, $V_{p'0}$ and $V_{p'p'}$
are vertex operators for open strings in the $0p'$, $p'0$, and $p'p'$
sectors respectively. 
The amplitude of this diagram is proportional to $g_{YM} \propto g_s^{1/2}$
\footnote{The disk diagram
has a normalization of $g_s^{-1}$. Each open string vertex operator insertion
carries a factor of $g_s^{1/2}$. Hence the overall amplitude of such
diagram is proportional  to $g_s^{1/2}$.}.
Higher order couplings between D-matter and the Standard Model fields are
suppressed by the higher power of $g_s$. Therefore, from power
counting, the S-matrix   describing the interactions of 
D-matter has an expansion in terms of $g_s$.

However, we should be more careful in writing down the effective coupling. From  
power counting, the amplitude describing the interaction of D-matter with
Standard Model fields
(whose lowest order contribution is shown in Fig.~3) 
has an expansion 
\begin{equation}
g_s^{\frac{1}{2}} F(s,t, \alpha') + g_s G(s,t, \alpha')+ ... 
\end{equation}
We see that the effective coupling extracted from this expression is
$(g_s^{\frac{1}{2}} F(s,t, \alpha') +...) \bar{\Psi} A^{\mu} \Psi$. In
other words, there is a ``form factor'' which 
effectively  ``renormalizes''  the coupling constant. In some
kinematical region, such as forward elastic scattering, we can safely
ignore this correction due to the low momentum exchange. In fact, in
this case, the deviation from a lowest order field theory result is
expected to be suppressed by powers of $q/M_S$ where $q$ is the
momentum exchange. However, in some other kinematical region of
interest, such as $s$-channel annihilation, the correction is not
expected to be small due to the fact that the energy involved exceeds
the string scale. In this regime, we expect corrections of stringy
nature to become significant.  
The question is then the magnitude of such corrections. 

\begin{center}
\begin{figure}[h]
\centering
\epsfxsize=6cm
\epsfbox{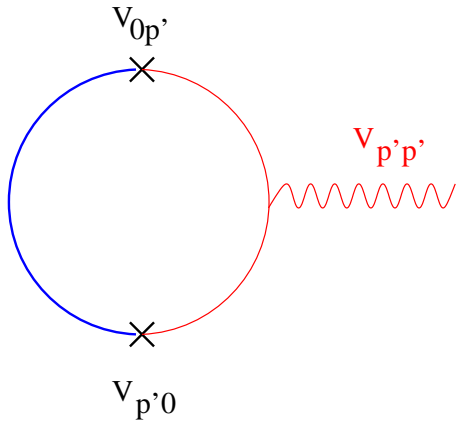}
\caption{At lowest order, the interaction of D-matter with the
Standard Model fields is described 
by a disk diagram with the insertion of open string vertex operators.}
\end{figure}
\end{center}

The techniques for calculating the
full annihilation cross section 
(i.e., including the stringy corrections)
of D-branes have not been fully developed.
Nevertheless,
we can estimate the size of the stringy 
corrections (i.e., $\alpha^{\prime}$ effects) by studying 
this annihilation process
in the dual heterotic
string picture. 
According to the Type I - heterotic duality \cite{Polchinski:1995df}, 
the stable D-particle in Type I string theory (an example of D-matter)
is dual to a massive heterotic string state \cite{Sen}.
In the dual heterotic string picture, the annihilation process
of interest involves only
perturbative string states and so the corresponding amplitude
can be calculated using standard techniques of
string perturbation theory.
The details of this heterotic string amplitude
calculation 
are given in the Appendix. 
The string amplitude can then be compared with the field theory result
(which corresponds to $\alpha' \rightarrow 0$).
As shown in Fig.~5,
the $\alpha'$ corrections over a wide range of energies
are of the same order as the field
theory result and so they cannot be neglected.

Hence, we expect that the D-matter gauge coupling would have an
${\mathcal{O}} (1)$ correction from the form factor. This correction
should be small when we consider elastic scattering with small
momentum exchange (small comparing to string scale). The correction
would be important when we consider $s$-channel
annihilations. However, we do not expect order of magnitude
enhancement or suppression. After all, the energy scale involved is
only slightly  above the string scale and so the corrections
only come from the lowest few string resonances. 
\begin{table}
\begin{tabular}{|c|c|c|}
\hline
& \qquad {'t Hooft-Polyakov Monopole} \qquad & D-Matter  \\
\hline
& & \\
Mass & $\frac{< \phi >}{g_{YM}} \sim \frac{M_X}{g_{YM}^2}$ & 
$\frac{M_s}{g_s} \sim \frac{M_s}{g_{YM}^2} $ \\
& & \\
\hline && \\
(size)$^{-1}$ & $~~~~\lambda < \phi>$ & $g_s^{\alpha} M_s$ \\
&  $g_{YM} < \phi >$ & \\
& &
\qquad $\alpha = \left\{ \begin{array}{ll} -1/3 & \mbox{brane-probe} \\
0 & \mbox{string-probe}
\qquad
\end{array}
\right.$
\\
& & \\
\hline &&  \\
Interaction & $\propto \mu_m = \frac{n}{g_{YM}}$ where $n \in {\bf Z}$
& $\propto g_{YM}$ \\ 
& & \\
\hline
\end{tabular}
\label{monopole}
\caption{Comparison between
't Hooft-Polyakov monopole and the D-Matter discussed in this section. Here, $<\phi>$ denotes the
vacuum expectation  value of the scalar field $\phi$ of the monopole
configuration,  $\lambda$ is its coupling constant, and $M_X$ is the symmetry
breaking scale. The size of D-matter depends on the probe since
D-branes can probe smaller distances than fundamental strings
\cite{Shenker,DKPS}.} 
\end{table}

  It is interesting to
  compare the properties of such D-matter with that of a 
't Hooft-Polyakov  monopole in spontaneously broken gauge  theory
which is also a non-perturbative object. 
They have similar properties as far as their masses and sizes go.
However, the 't Hooft Polyakov monopole carries a magnetic charge
$\mu_m \propto 1/g_{YM}$ 
under the unbroken $U(1)$. Therefore, unlike the D-matter considered here, the long range gauge
interactions between 't Hooft Polyakov monopoles are non-perturbative
\footnote{Note however that some D-matter states could act as magnetic point sources of
the worldvolume gauge field on some higher-dimensional branes, in which case
their properties are similar to that of 't Hooft Polyakov magnetic monopoles. For example, a
D1-brane ending on two D3-branes corresponds to a magnetic monopole of the
$SU(2)$ gauge theory on the D3-brane worldvolume. This configuration is S-dual to
a fundamental string ending on the D3-branes. The power counting of $g_s$ for
D-matter interactions can
be obtained by taking $g_s \rightarrow 1/g_s$ of the S-dual amplitudes.}.
A comparison between 't Hooft Polyakov monopole and the D-matter discussed here
is summarized
in Table I.

\section{Phenomenology of D-Matter}\label{phenomenology}

In this section, we study some phenomenological features of
D-matter. The spectrum and quantum numbers  of the
D-matter states are model-dependent. 
Instead of studying their properties in
the context of a specific model, we focus on the following two
generic features. 

\subsection{Cold Dark Matter Candidate}

As discussed above, the LDPs are stable and so
they could be candidates for the cold dark matter of our
universe. 
Their properties as dark matter candidates depend on the value of the string scale. 
In the
  conventional high string scale scenario ($M_S \sim 10^{16}$ GeV),
 the LDPs have masses greater than $10^{16}$
  GeV. They 
  are presumably removed by inflation (just as the GUT monopoles)
    and would not be produced
  significantly during reheating. Therefore,
  we consider the following two scenarios with intermediate and low string
  scales, respectively.

\begin{enumerate}
\item {\it Intermediate string scale: $M_s \sim 10^{11} -10^{12}$ GeV}

In this scenario, the mass of the LDP will be of the order of $10^{11} -
10^{12}$ GeV. It therefore fits in the category of 
wimpzillas, i.e., super-massive wimps. Since their
masses are higher than the reheating temperature, a  key question
is whether they can be produced in significant amount to
account for the cold dark matter in the universe. This question has
been studied in a series of papers
\cite{Chang:1996vw,Chung:1998zb,Chung:1998ua,Chung:1998is,Chung:1998rq,Chung:1998bt,Kolb:1998ki}.  
Several scenarios have been proposed to
produce them including preheating , non-adiabatic expansion of
the universe, and normal reheating. Each
of these mechanisms has its own special properties. It is interesting that
sufficient number of the superheavy LDPs can be produced through 
normal reheating.  From Ref.~\cite{Chung:1998rq}, 
the relic abundance of such a state 
produced during reheating is 
\begin{equation}
\Omega_D h^2 = M_D^2 <\sigma v> \left( \frac{g_*}{200}\right)^{-3/2}
\left(\frac{2000 T_{RH} }{M_D} \right)^{7}.
\end{equation}

The absence of the naively expected $\exp (-M_D/T_{RH})$ suppression
is due to the fact that during the reheating process, the maximum
temperature is much higher than the reheating temperature.  Using the
expression for the production cross section, we see that states as
heavy as $10^3 T_{RH}$ could be produced in an appropriate amount to
account for the cold dark matter in the universe. 

On the other hand, if the LDPs only couple to the light
states (or inflaton) with much weaker (or not at all) couplings (such
as gravitational strength couplings), they could be produced in a
``non-adiabatic'' expansion stage at the end of  inflation
\cite{Chung:1998zb}.
 
An interesting consequence of super-heavy dark matter is that
their annihilation could give rise to ultra-high energy cosmic rays
(UHECRs) with energies $> 10^{20}$ eV. The possibility of decaying
super heavy dark matter within galactic halo as a source of UHECR
events\cite{agasa,hires}  
which exceed the GZK cutoff \cite{Greisen:1966jv,Zatsepin:1966jv} has
been studied \cite{xdecay,Sarkar:talk}. Although the rate of UHECR
from the annihilation of D-matter is expected to be smaller than
the measured value (without
special assumptions about their couplings),  it is
possible that they may account for some  of the events provided that
some special features of the local density of dark 
matter are satisfied. If this is indeed 
the source of the UHECRs, another interesting correlated signal will be
an enhanced possibility of detecting high energy cosmic ray neutrinos
in Amanda II and IceCube \cite{UHECRnu}.

\item {\it Low string scale: $M_s \sim $ TeV}

Such a low string scale implies that our stable D-matter has a mass
around TeV as well. Suppose they interact with each other and the
light degree of freedoms in the thermal bath with the strength of
gauge interactions, then they have many similarities with
the well studied thermal relics (such as the LSP in MSSM). They will
be produced and in thermal equilibrium with the thermal bath during
reheating. Then as the universe expands, they will follow a
standard freeze out procedure. A key quantity which determines the relic abundance is
their annihilation cross section $<\sigma v>_{A}$. Since the D-matter
behaves as a particle in 4-dimensions, we could write down an
effective field theory for its coupling to the gauge bosons. In
particular, there will be a term of the form $ g_D \bar{D} D A$ (we
suppress the Lorentz structure which may contain gamma matrix or
spacetime derivative). The coupling constant $g_D$ should be
determined by a string theoretic calculation (generically, it has the
strength of a normal gauge coupling). The thermal relic abundance as a
result of the freeze out process is 
\begin{equation}
\Omega_D h^2 \sim \frac{3 \times 10^{-6}}{\alpha_D^2} \left( \frac{M_D}{100
{\mathtt{GeV}} } \right)^2.
\end{equation}
From this we see that for a gauge coupling strength interaction, we
can have a relic abundance appropriate for the CDM. Notice that this
case is very similar to the well studied LSP cold dark matter in the context of MSSM.

Let us discuss the prospects of detecting the LDPs. Direction
detection of dark matter\cite{Griest:kj}  relies on  
the elastic scattering  of dark matter off the nuclei within the
detector. As discussed in Section~\ref{interaction}, we do not expect
a sizable modification of the effective gauge coupling in this
channel. Therefore, we should be able to calculate their rate
reliably. However, there is  
a subtlety we need to keep in mind. The dark matter-nucleon
cross-section is sensitive to the spin of the dark matter particle
\cite{Griest:kj}.  For example, in MSSM, a neutralino
(which is a Majorana fermion) and a sneutrino (which is a boson) have
very different direct detection cross-sections. 
Similarly, D-matter states which are bosonic and fermionic respectively
\footnote{D-matter states could be bosonic or fermionic. They
  correspond to the bosonic and fermionic 
zero modes of the stable D-branes respectively.} will have different
  dark-matter-nucleon cross-sections.   

Indirect detection of the dark matter relies on detecting the cosmic
rays resulting from their annihilation within the galactic halo. As
discussed in Section~\ref{interaction}, we expect the annihilation
cross section to be modified with respect to the field theory result. 
This may have observable effects in high energy cosmic rays such
as an excess (or deficiency) of cosmic ray flux.

\end{enumerate}

\subsection{Excited States and Mass Splittings}

Another interesting feature of D-matter is the mass spectrum of its
excited states. 
A tower of excited states can be constructed by
dressing the D-matter with excitations of open strings attached to it.
These D-matter excited states could be a charged state or a singlet under
the gauge symmetries of the background D-branes
depending on whether one or both endpoints of the open strings 
are attached to the D-matter  \footnote{Notice that charge
conservation with respect to the unbroken gauge group  on the worldvolume of 
D-matter has to be
satisfied if only one endpoint of the open strings is attached to it. In
principle, the gauge symmetry on the world volume of  D-matter
could be broken in a realistic model so  this type of
requirement would not impose a restriction on the possible types of
excited states}. Thus, these two types of excited states could lead to
different patterns of decay modes. 
An example of such an excited state is shown in Fig.~2.
Schematically, an excited D-matter state can be written as 
\begin{equation}
|D_0> \times | {\cal{V}}>,
\end{equation} 
where $|D_0>$ is the ``bare'' D-matter and $ {\cal{V}}$ is
the vertex operator of some excited open string state. The mass of
such an excited state is given by  \footnote{At a more technical level, 
the zeroth mode of the Virasoro generator of an excited D-particle state is
$L_0=L^D_0 + L_0^{\cal V}$.  Requiring $L_0$ annihilate the D-matter state
will give us the mass formula.}:
\begin{equation}
M_{D^*}^2 = M_{D_0}^2 + n M_s^2,
\end{equation}
where $M_{D_0} \sim M_s/g_s$. This is because
a string resonance contributes to the
{\it square} of the mass of a state by an amount of $n M_s^2$
where $n$ is the resonance level.

\begin{figure}[h!]
\centering
\includegraphics[scale=0.7,angle=270]{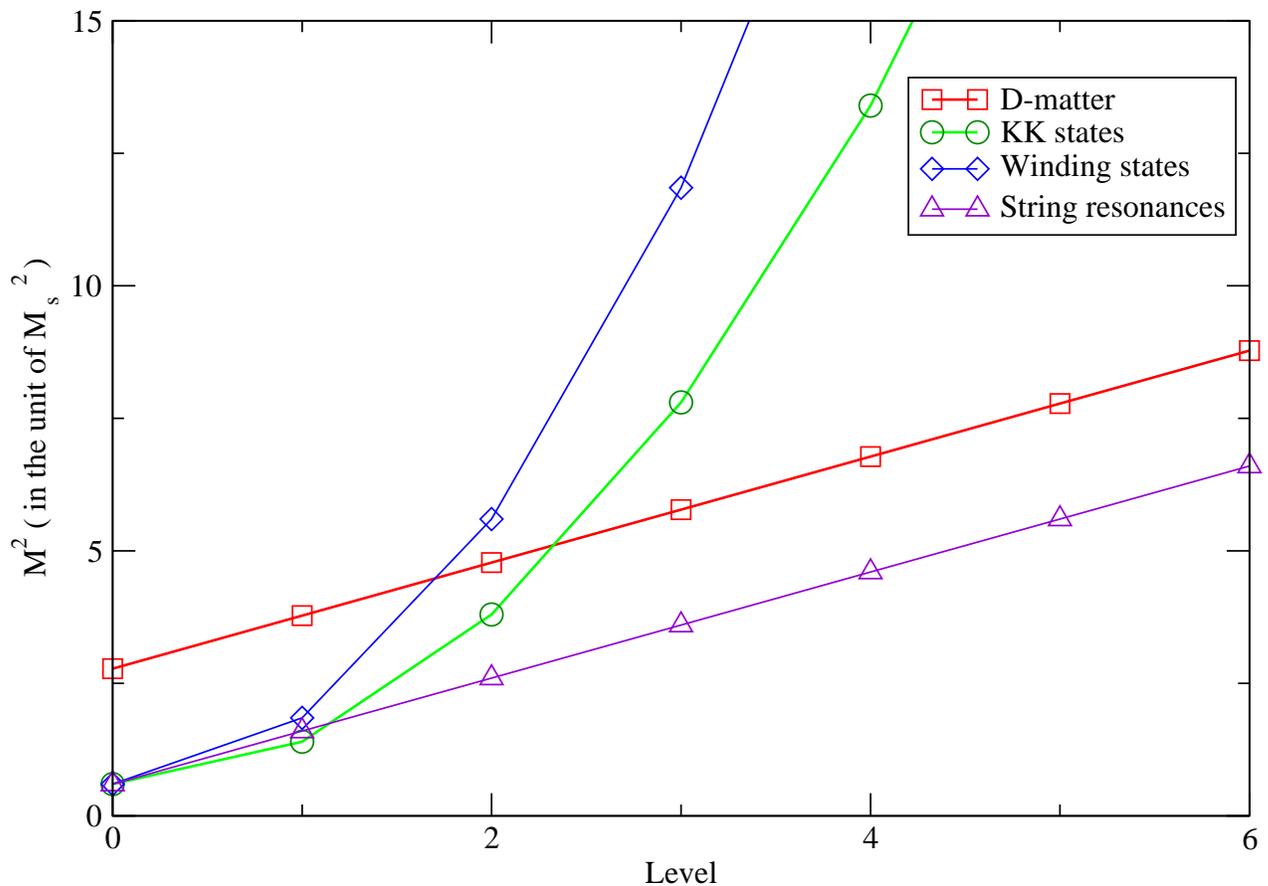}
\label{tower}
\caption{Different towers of excited states. The masses are
given in unit of $M_S$. For illustrative purpose, we choose a
representative set of parameters: the compactification scale
$M_C \sim R^{-1}$ is taken to be $0.8 M_s$, and the string coupling $g_s
\sim 0.6$. With some symmetry breaking scenario in mind, we choose a
zero point energy for the zero modes to be $g_s M_s$. }
\end{figure}

Let us compare this tower of excited states with the familiar
Kaluza-Klein (KK) modes, winding modes, and string resonances:
\begin{eqnarray}
M^2_{KK} &=& g_s M_s^2 + n^2 M_C^2, \nonumber \\
M^2_{\mbox{\small winding}} &=& g_s M_s^2 + n^2 \frac{M_s^4}{M_C^2} \nonumber \\
M^2_{\mbox{\small resonances}} &=& g_s M_s^2 + n M_s^2 \nonumber \\
M^{2}_{D^*} &=& \frac{M_s^2}{g_s^2} + n M_s^2, 
\end{eqnarray}
where $M_C$ is the compactification scale.
The zero point masses of the KK modes, winding modes, and string resonances  are
taken to be $g_s M_s^2 \sim
g_{YM}^2 v^2$, which
presumably come from some spontaneous symmetry breaking. 
A sample spectrum of the various towers of excited
states is shown in Fig.~4. 
In the case of a low string scale scenario, these different
towers of new particles  
(whose mass spectra have distinct patterns) can have interesting
observable signatures at colliders. 
Note that the open string states attached to the D-matter can 
in general carry
momentum and/or winding quantum numbers as well, 
but the details depend
on the origin of the D-matter and how the Standard Model is embedded
\footnote{Whether the open string states carry momentum/winding numbers depends
on whether the corresponding compact dimension is along or transverse to the
worldvolumes of D-matter and the Standard Model (SM) branes.
If a compact dimension is along the worldvolumes
of both the D-matter and the SM branes
(Neumann-Neumann boundary condition), the open string states
can carry momentum but not winding numbers.
Conversely, if a compact dimension is transverse to both
the D-matter and the SM branes (Dirichlet-Dirichlet
boundary condition), the open string states
can carry winding but not momentum numbers.
Finally, if a compact dimension is transverse to either the D-matter or the SM branes but not both
(Dirichlet-Neumann boundary condition),
the open string states do not
carry any momentum or winding numbers.}.
For illustrative purpose, we consider the case
where the D-matter is a stable D-particle (e.g., in Type I string theory) and the Standard Model 
is embedded on a set of D9-branes.
In this case, 
the open string states have no momentum or winding numbers.
In general, the possibility of momentum/winding quantum numbers
will give rise to an even richer spectrum than the one shown in 
Fig.~4.

For simplicity, we have taken the zero point mass of the
D-matter state to be $m_D \sim {M_s}/{g_s}$.
This is based on the crude assumption that the volume of the cycle $V_p$ that the
stable D-brane wraps around is of the order of $M_s^{-1}$. 
In general, $m_D \sim M_s^{p+1} V_p/g_s$.
Therefore, the zero point mass of D-matter 
carries some useful information about the compact 
dimensions, which may allow us to distinguish between
isospectral manifolds (different manifolds which nonetheless have the same
spectrum) that are indiscernible by the KK and winding modes
\cite{Rabadan:2002wb}.

\section{Discussions and Conclusion}\label{conclusion}

In this paper, we study the phenomenology of D-matter, i.e.,
particle-like states originated from 
D-branes whose spatial dimensions are all wrapped around the compact space.
An interesting feature of D-matter is that although they are
non-perturbative objects, 
they could have perturbative couplings with each other and with the
Standard Model fields. 
Therefore, the D-matter states are weakly interacting and the lightest
stable D-matter states (whose stability is due to integral or torsion
charges  
they carry) are candidates for wimps or wimpzillas depending on the
string scale. In the case of a low string scale, there are also
potentially rich collider phenomenologies. Exploring the mass pattern of
the lower lying D-matter states could reveal additional information about
the structure of the compact dimensions.For illustrative purposes, we
consider the simple case where all the Standard Model 
particles are localized on a single set of $Dp'$-branes. It would be
interesting to explore the phenomenology of D-matter in
more realistic brane world models such as models with background intersecting 
D-branes \cite{CSU1,CSU2,CSU3,CIS,Berkooz,Berlin1,Sagnotti,Madrid1,Berlin2,Madrid2,Berlin3}.
In these models, the gauge bosons and matter fields are localized
differently in the extra dimensions (e.g., in different 
codimensions) and hence they will couple with different strengths to
the D-matter.

Another interesting direction motivated by this work is
to compute the spectra of stable D-branes (and their corresponding K-theories) 
in some realistic brane world models, e.g.,
certain four-dimensional $N=1$ supersymmetric orientifold models
whose 
perturbative string 
spectra contain the Standard Model or Grand Unified Theories
\cite{CSU1,CSU2,CSU3,CIS}.  
The K-theories which classify the D-brane spectra
have been computed for some concrete orientifold 
models, e.g., orientifolds with $N=2$ supersymmetries
(in a four-dimensional sense) in  \cite{Quiroz,Braun}.
It would be interesting to extend these analyses to
models with $N=1$ supersymmetry
where chiral fermions can appear.
Such analyses could also help to better understand the types of defects 
(from a four-dimensional perspective) that 
are formed toward the end of brane inflation \cite{DvaliTye},
besides the cosmic strings that are generically produced  \cite{Jones,Sarangi,ShiuTyeWasserman,Shiu,Pogosian}.
Therefore, in addition to the conditions one often imposes
(such as appropriate gauge groups,
three generations of chiral matter, etc)  in estimating the statistics of 
realistic string theory
vacua \cite{Douglas:2003um}, the
cosmological constraints on the existence of stable defects
(such as cosmic strings and domain walls) 
may further restrict the space of viable string models.

\appendix

\section{Four-Point Amplitudes in Heterotic String Dual}

\noindent The stable non-BPS D-particles in Type I are dual to the first
massive string states
of the $SO(32)$ heterotic string \cite{Sen}. The vertex operators of
such massive string states in the R-sector (spacetime fermions) in the -1/2 picture
are given by: 
\begin{equation}\label{massive}
V^{-1/2} (z, \overline{z}) = e^{-\phi (z)/2} e_{\mu} u_{\alpha} i \partial_{z} X^{\mu} \theta^{\alpha} (z)
e^{i k \cdot X(z,\overline{z})}
C_{K} e^{i \frac{K}{\sqrt{2 \alpha}} \cdot \tilde{X}(\overline{z})} 
\end{equation}
and
\begin{equation}
V^{-1/2} (z, \overline{z}) = e^{-\phi (z)/2} e_{\mu} u_{ \alpha} i \partial_{z} \psi^{\mu} 
\theta^{\prime \alpha} (z)
e^{i k \cdot X(z,\overline{z})}
C_{K} e^{i \frac{K}{\sqrt{2 \alpha}} \cdot \tilde{X}(\overline{z})} 
\end{equation}
where $k^2= -M^2 = - 4/\alpha'$ and $K^2=4$. Here $\theta_{\alpha}$, $\theta^{\prime}_{\alpha}$ are the ten-dimensional
spin fields with opposite chirality (because of the
GSO projection and the fact that $\psi$ changes the fermion number by 1), $e_{\mu}$ is the polarization (the physical state condition gives $e \cdot k =0$), $X^{\mu} (z,\overline{z})$ for $\mu =0,\dots,9$
are the 10-dimensional
coordinates for both left- and right-movers, $\tilde{X}^I (\overline{z})$ for $I=1,\dots,16$
are the 16-dimensional
internal coordinates corresponding to the $SO(32)$ lattice, $C_K$ is a cocycle factor,
and $\phi$ is the boson arising from the bosonization of the superconformal ghosts. 
Using the fact
that the conformal dimension of $e^{q \phi}$ is $h=-q(q+2)/2$, it is easy to see that
the above vertex operators have conformal dimensions $(h,\overline{h})=(1,1)$. 
We take $K = (\pm \frac{1}{2},\pm \frac{1}{2},\dots,\pm \frac{1}{2})$ with an even number of $+$
sign which clearly gives $K^2=4$. These massive states indeed transform in the spinor rep.
of $SO(32)$.
In addition to the $SO(32)$ quantum numbers, the two types of vertex operators above together contribute to the {\bf 128} representation of the
Lorentz group.
For simplicity, we focus on the first type of vertex operators in Eqn.(\ref{massive})
in the calculation of  four-point amplitudes.

The scattering of two D0-branes to two fundamental open string states corresponds to the scattering of two massive string states to two massless states
in the heterotic dual.
The tree-level diagram has $\phi$-ghost number -2, and so the four-point amplitude
can be computed with all external states in the -1/2 picture. In other words, the
scattering amplitude of two massive spacetime fermions in the spinor reps of SO(32) into
two gauginos:
\begin{equation}
A_4 \sim {\cal C}_0 \hat{N}^4 < c \overline{c}  V_1^{-1/2} (z_1,\overline{z}_1) 
 V_2^{-1/2} (z_2,\overline{z}_2) 
 V_3^{-1/2} (z_3,\overline{z}_3) 
 V_4^{-1/2} (z_4,\overline{z}_4)  >
\end{equation}
where ${\cal C}_0$ and $\hat{N}$ are normalization factors defined in  \cite{Gallot},
$V_1$ and $V_2$ are the vertex operators for the massive string states in the spinor rep.
as in Eqn.~(\ref{massive})
above, $V_3$ and $V_4$ are vertex operators for the gauginos as follows:
\begin{equation}
V^{-1/2}  (z, \overline{z}) = e^{-\phi (z)/2} u_{\alpha} \theta^{\alpha} (z)
e^{i k \cdot X(z,\overline{z})} {\cal O} (\overline{z})
\end{equation}
with $k^2=0$. ${\cal O} (\overline{z})$ is the operator for the $SO(32)$ degrees of 
freedom, which is equal to $\overline{\partial} \tilde{X}^I (\overline{z})$ for the
Cartan generators and
$C_{K} e^{i \frac{K}{\sqrt{2 \alpha}} \tilde{X}(\overline{z})}$ with $K^2=2$
for the remaining 480 generators. We can take
$K$ of the form $(0,\dots,\pm 1,0, \dots,0,\pm 1,0,\dots,0)$.
In either case, $\overline{h}_{\cal O} = 1$, so it is easy to see that $(h, \overline{h})=(1,1)$.  

Let ${\cal V}_{\alpha} (z) = e^{-\phi/2} \theta_{\alpha} (z)$,
the relevant correlation function (see, e.g., eqn (12.4.19) of \cite{Polchinski}) is:
\begin{equation}
< {\cal V}_{\alpha}  (z_1) 
{\cal V}_{\beta}  (z_2) 
{\cal V}_{\gamma}  (z_3) 
{\cal V}_{\delta}  (z_4) >
= \frac{ (C \Gamma^{\mu})_{\alpha \beta} (C \Gamma_{\mu})_{\gamma \delta}}{2 z_{12}z_{23}
z_{24} z_{34}}
+ \frac{ (C \Gamma^{\mu})_{\alpha \gamma} (C \Gamma_{\mu})_{\delta \beta}}{2 z_{13}z_{34}
z_{32} z_{42}}
+ \frac{ (C \Gamma^{\mu})_{\alpha \delta} (C \Gamma_{\mu})_{\beta \gamma}}{2 z_{14}z_{42}
z_{43} z_{23}}
\end{equation}
where $C$ is the charge conjugation operator.

The correlation function of the bosonic fields gives
\begin{equation}
< i \partial_z X^{\mu} e^{i k_1 \cdot X} (z_1,\overline{z}_1)
i \partial_z X^{\nu} e^{i k_2 \cdot X} (z_2, \overline{z}_2)
e^{i k_3 \cdot X} (z_3,\overline{z}_3) e^{i k_4 \cdot X} (z_4, \overline{z}_4) >
= \prod_{i<j} |z_{ij}|^{\alpha' k_i \cdot k_j} V^{\mu \nu}
\end{equation}
where
\begin{equation}
V^{\mu \nu} = \frac{\alpha' \eta^{\mu \nu}}{2 z_{12}^2}
+ \frac{\alpha^{\prime 2}}{4} \sum_{i \not= 1,j \not= 2}
\frac{k_i^{\mu} k_j^{\nu}}{z_{1i} z_{2j}}
\end{equation}
Note that in the three-point function of massless states, one can use transversality 
($e \cdot k =0$) to
simplify the expression (see eqn 12.4.12 of \cite{Polchinski}).
The transversality condition comes from
the physical state conditions: $L_{m} -a \delta_{m,0} = G_{r} = 0$ for $m \geq 0$, $r \geq 0$.
It is easy to check that the physical state condition also gives $e \cdot k=0$ for the
massive states we consider here.

For the $SO(32)$ degrees of freedom:
\begin{equation}
< e^{i \frac{K_1}{\sqrt{2 \alpha'}} \cdot \tilde{X} (\overline{z}_1)}  e^{i \frac{K_2}{\sqrt{2 \alpha'}} \cdot \tilde{X}  (\overline{z}_2)}
e^{i \frac{K_3}{\sqrt{2 \alpha'}} \cdot \tilde{X} (\overline{z}_3)}
e^{i \frac{K_4}{\sqrt{2 \alpha'}} \cdot \tilde{X} (\overline{z}_4)} >  = \prod_{i<j} \overline{z}_{ij} ^{K_i \cdot K_j}
\end{equation}

To calculate the four-point amplitude,
we can set $z_1=0$, $z_2 = x$, $z_3=1$, $z_4=\infty$, and then sum over all permutations.
The momentum dependence can be simplified as usual with the Mandelstam $s,t,u$
variables.

Let us focus on one of the amplitudes (the ordering 1234):
\begin{equation}
<c \overline{c} (z_1) c \overline{c} (z_3) c \overline{c} (z_4)> =
(z_4 \overline{z}_4 )^2 
\end{equation}
\begin{equation}
< {\cal V}_{\alpha}  (z_1) 
{\cal V}_{\beta}  (z_2) 
{\cal V}_{\gamma}  (z_3) 
{\cal V}_{\delta}  (z_4) >
= \frac{(\overline{u}_1 \cdot u_2)(\overline{u}_3 \cdot u_4)}{2 x(1-x) z_4^2}
+ \frac{(\overline{u}_1 \cdot u_3)(\overline{u}_4 \cdot u_2)}{2 (1-x) z_4^2}
\end{equation}
\begin{equation}
V^{\mu \nu} = \frac{\alpha' \eta^{\mu \nu}}{2 x^2}
+ \frac{\alpha'}{4} \left( -\frac{k^{\mu}_2 k^{\nu}_1}{x^2} -
\frac{k_3^{\mu} k_1^{\nu}}{x} 
+\frac{k^{\mu}_2 k^{\nu}_3}{x(1-x)} + \frac{k^{\mu}_3 k^{\nu}_3}{1-x} \right)
\end{equation}
\begin{eqnarray}
\prod_{i<j} |z_{ij}|^{\alpha' k_i \cdot k_j} &=& |z_4|^{-\alpha'
  k_4^2} (x \overline{x})^{\alpha' k_1 \cdot k_2/2}
\left[(1-x)(1-\overline{x}) \right]^{\alpha' k_2 \cdot k_3/2}
\nonumber \\ 
&=& (x \overline{x})^{-\alpha' s/4} \left[ (1-x) (1-\overline{x})
  \right]^{-\alpha' u/4} 
\end{eqnarray}
\begin{equation}
\prod_{i<j} \overline{z}_{ij}^{K_i \cdot K_j} =
\overline{z_4}^{-K_4^2} \overline{x}^{K_1 \cdot K_2} 
(1-\overline{x})^{K_2 \cdot K_3}
=\overline{z}_4^{-2} \overline{x}^{-S/2} (1- \overline{x})^{-U/2}
\end{equation}
Note that the powers of $z_4$ and $\overline{z}_4$ are canceled. There are many
terms in the final expression. Let's focus on the term proportional to
$\eta^{\mu \nu}$: 
\begin{equation}
C_0 \hat{N}^4 (e_1 \cdot e_2) \int d^2 x x^{-\alpha' s/4-1}
 \overline{x}^{-\alpha' s/4 - S/2-2} 
 (1-x)^{-\alpha' u/4} (1-\overline{x})^{-\alpha' u/4 - U/2+1}
\end{equation}

We can express the amplitude in terms of $\Gamma$-functions and
calculate the correction to the tree-level field theory result.  
The integral we have to do is
\begin{eqnarray}
\int_C d^2z z^{-\lambda_s-1} \bar{z}^{-\lambda_s-S/2-2}
(1-z)^{-\lambda_u} (1-\bar{z})^{-\lambda_u-U/2+1},
\label{integral0}
\end{eqnarray}
where $\lambda_s=\alpha' s /4 $, $\lambda_u=\alpha' u/4$. We will suppress
the $SO(32)$ structure. Consider the following
choice of internal momenta
\begin{eqnarray}
K_1 &=& - K_2 = (+\frac{1}{2},....,+\frac{1}{2}), \nonumber \\
K_3 &=& - K_4 = (0,...0,+1,0,...,0,+1,0,...).
\end{eqnarray}
With this choice, we have $S=0$, $U=-4$. The integral reduces to 
\begin{eqnarray}
\int_C d^2z z^{-\lambda_s-1} \bar{z}^{-\lambda_s-2}
(1-z)^{-\lambda_u} (1-\bar{z})^{-\lambda_u+3}.
\label{integral}.
\end{eqnarray}
We can choose the following space-time momenta:
\begin{eqnarray}
k_1 &=& (E,0,...0,k) \nonumber \\
k_2 &=& (E,0,...0,-k) \nonumber \\
k_3 &=& (-E,0,...-E \cos \theta, -E \sin \theta) \nonumber \\
k_3 &=& (-E,0,...E \cos \theta, E \sin \theta). 
\end{eqnarray}
For example, at fixed angle $\theta=\pi/2$, we have 
\begin{eqnarray}
s &=& 4E^2 \nonumber \\
u &=& -2E^2+m^2,
\end{eqnarray}
where $m^2=4/\alpha'$.

The $s=0$ pole could be extracted as follows. First, we expand the
integrand of Eq.~(\ref{integral}) around $z=0$ and obtain
\begin{eqnarray} 
I &\sim& \int_C d^2 z ~\left[ (z \bar{z})^{-\lambda_s-1} \bar{z}^{-1} + (z
\bar{z})^{-\lambda_s-1} \bar{z}^{-1} z + ... \right. \nonumber \\
&& \qquad \qquad + \left. (\lambda_u-3) (z \bar{z})^{-\lambda_s-1} + ... \right]
\end{eqnarray}
The terms such as the ones displayed on the first line do not contribute
to the pole. Therefore, the $s=0$ pole is from the second line:
\begin{equation}
-\frac{1}{2}(\lambda_u-3)\frac{1}{\lambda_s}.
\end{equation}
The full amplitude is 
\begin{eqnarray}
{\mathcal{A}} &\propto& 2 \pi \times \frac{\Gamma
  (\lambda_s+\lambda_u-2) \Gamma (-\lambda_s) \Gamma (1-\lambda_u) }{
  \Gamma (\lambda_u -3) \Gamma (\lambda_s +2) \Gamma (1- \lambda_s - \lambda_u)}.
\end{eqnarray}
We can use the $\Gamma$-function identity $\Gamma(x) \Gamma (1-x) = \pi/
\sin (\pi x)$ to rewrite this amplitude as
\begin{eqnarray}
{\mathcal{A}} &\propto& 2 \times \frac{\Gamma
  (\lambda_s+\lambda_u-2) \Gamma (1-\lambda_u)  \Gamma (4-\lambda_u)
  \Gamma (\lambda_s+\lambda_u) }{ 
   \Gamma (\lambda_s+2) \Gamma (\lambda_s )} \nonumber \\
&\times& \frac{\sin [\pi (4-\lambda_u)] \sin
  [\pi(\lambda_s+\lambda_u)]}{\sin [\pi \lambda_s]} \times \left(
  \frac{-1}{\lambda_s}\right). 
\end{eqnarray}
From this we clearly see the closed string poles at $\lambda_s =
4, 5, 6.....$, which correspond to the production of on-shell
states in this channel. 

\begin{figure}[h!]
\centering
\includegraphics[scale=0.5,angle=270]{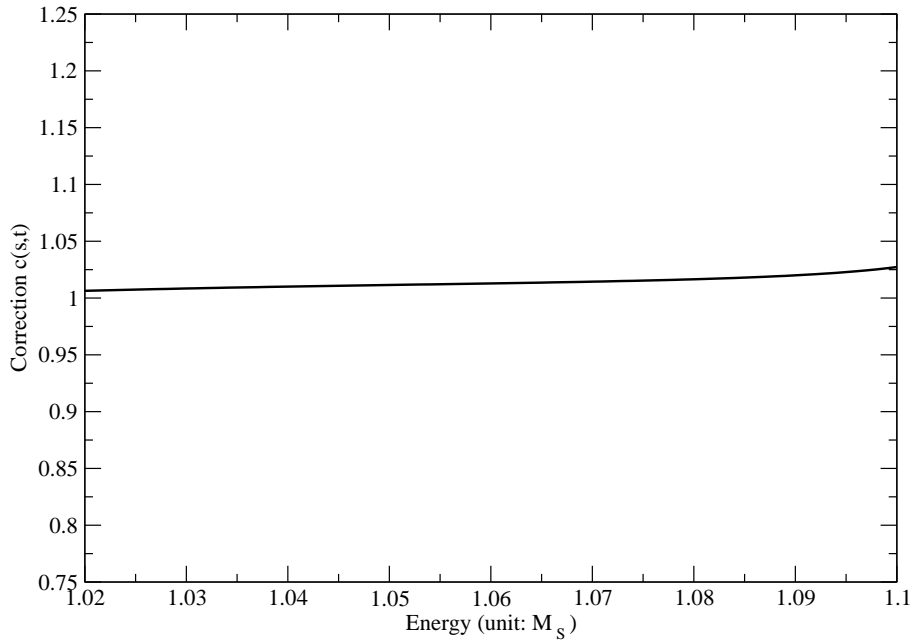}
\label{correction}
\caption{The corrections of the form factor. }
\end{figure}
We could write the amplitude in the form of ${\mathcal{A}}=  f(u,t)/s + g(s,t,u,
\alpha')$, where $g(s,t,u,\alpha')$ is the stringy correction to the
zero mass gauge boson exchange in field theory. We could define a
measure of the stringy corrections as follows: 
\begin{equation}
\frac{g(s,t,u,\alpha')}{f(u,t)/s} = \frac{{\mathcal{A}}}{f(u,t)/s} -1,
\end{equation}
This quantity as a function of energy $E$ is shown in Fig.~5 from which
we see that the correction is of order one in a wide range of energies.

\acknowledgments

We thank Daniel Chung, Michael Douglas, Aki Hashimoto, Nemanja Kaloper, Gordy Kane,
Fernando Marchesano, Rob Myers, Asad Naqvi, Raul Rabadan, Koenraad Schalm,
Bogdan Stefanski,
Henry Tye, and Angel Uranga for discussions. LW would like to thank
Aspen Center for Physics for hospitality where part of this work was
completed. 
The work of GS was supported in part by funds from the University of Wisconsin.
The work of LW was supported in part by a DOE grant No. DE-FG-02-95ER40896
and the Wisconsin Alumni Research Foundation.

\end{document}